# Cracking in polymer substrates for flexible devices and its mitigation

Anush Ranka[1], Madhuja Layek[1], Sayaka Kochiyama[1], Cristina López-Pernia[1],
Alicia M. Chandler[1], Conrad A. Kocoj[2,3], Erica Magliano[4], Aldo Di Carlo[4,5], Francesca Brunetti[4],
Peijun Guo[2,3], Subra Suresh[1], David C. Paine[1], Haneesh Kesari[1], and Nitin P. Padture[1*]

[1] School of Engineering, Brown University, Providence, RI 02912, USA.

[2] Department of Chemical and Environmental Engineering, Yale University, New Haven, CT 06520, USA.

[3] Energy Sciences Institute, Yale University, West Haven, CT 06516, USA.

[4] Centre for Hybrid and Organic Solar Energy, Department of Electronic Engineering, University of Rome Tor Vergata, Rome 00133, Italy.

[5] Instituto di Struttura della Materia, National Research Council, Rome 00133, Italy.

[*] Email: nitin_padture@brown.edu

**Mechanical reliability plays an outsized role in determining the durability of flexible electronic devices because of the significant mechanical stresses they can experience during manufacturing and operation. These devices are typically built on sheets comprising stiff thin-film electrodes on compliant polymer substrates, and it is generally assumed that the high-toughness substrates do not crack easily. Contrary to this widespread assumption, here we reveal severe, pervasive, and extensive cracking in the polymer substrates during bending of electrode/substrate sheets, which compromises the overall mechanical integrity of the entire device. The substrate-cracking phenomenon appears to be general, and it is driven by the amplified stress intensity factor caused by the elastic mismatch at the film/substrate interface. To mitigate this substrate cracking, an interlayer-engineering approach is designed and experimentally demonstrated. This approach is generic, and it is potentially applicable to myriad flexible electronic devices that utilize stiff films on compliant substrates, for improving their durability and reliability.**

There is growing interest in flexible electronic devices used in a wide range of applications such as:[1,2] energy harvesting and storage; displays and lighting; sensors and actuators; health-monitors and wearables; *etc*. By some estimates the global market for flexible electronics is expected to exceed 75 billon US dollars by 2032.[3] The foundation of these devices is the flexible substrate, which is typically made of polymers such as:[4] polyethylene naphthalate (PEN), polyethylene terephthalate (PET), polyimide (PI), polydimethylsiloxane (PDMS), *etc*. For many applications, such as organic photovoltaics (OPV),[5] perovskite solar cells (PSCs),[6] displays/lighting,[7] touchscreens,[8] electrochromics,[9] *etc*. the polymer substrate is coated with transparent-conducting oxide (TCO) thin film, which is typically made of:[10] indium-tin oxide (ITO), indium-zinc oxide (IZO), aluminum-zinc oxide (AZO), *etc*.

Since the mechanical reliability of flexible electronic devices built on these TCO/polymer sheets is so important, several mechanical-testing methods have been developed.[11,12] Among them, the bending test is the most commonly used to study the mechanical behavior and failure mechanisms in flexible electronic devices.[13] This test involves bending the device around a cylindrical mandrel of radius, $r$, subjecting the entire active part of the device to uniaxial tension (Fig. 1a).[13] Since the polymer substrates toughness is several orders of magnitude higher than that of the ceramic TCO thin film, the substrates are generally not expected to crack in bending. For example, the PEN has a specific effective work of fracture or SEWF ($w_e$) [14] of ~55,000 J·m⁻²,[15] whereas the equivalent toughness, $G_C$, of ITO is ~5 J·m⁻².[16]

**Bending tests of TCO/polymer sheets**

Here, commercially available ITO/PEN sheets, which are commonly used in flexible electronic devices, were subjected to bending tests. Figure 1b is a top-view optical micrograph of the ITO/PEN sheet in the bent state ($r = 7$ mm), where the externally applied uniaxial tensile strain at the convex surface and near-surface region is given by $\varepsilon_A = t_S/2r$,[13,17] is 0.0089 ($t_S$ is the substrate thickness, ~125 μm for PEN). (Supplementary Fig. 1 plots the calculated applied uniaxial tensile universal strain ($\varepsilon_A$), as a function of $r$, and the corresponding stress ($\sigma_A$) within the layers studied here. See Supplementary Note 1 regarding multilayers.[13] Any tensile or compressive internal residual stresses within the individual layers would augment or diminish, respectively, the overall stresses in the layers in the bent state. Supplementary Table 1 lists the thicknesses of the different layers used in this study and their relevant static mechanical properties.[15, 18-26]) Under this applied stress, typical array of 'channel' cracks running parallel to the bending axis within the brittle ITO film are observed in Fig. 1b. Such observations have been reported in the flexible-electronics literature on numerous occasions (see *e.g.* ref.[16]). As mentioned earlier, it is generally assumed that the polymer substrate does not crack due to its extremely high toughness and substantial thickness relative to the TCO thin film (*e.g.* ITO thickness is ~300 nm).[16] This purported assumption is supported by the observations in Fig. 1b where the 'channel' cracks in the ITO film do not appear to extend into the bare PEN substrate part without the ITO film (bottom region of the micrograph).

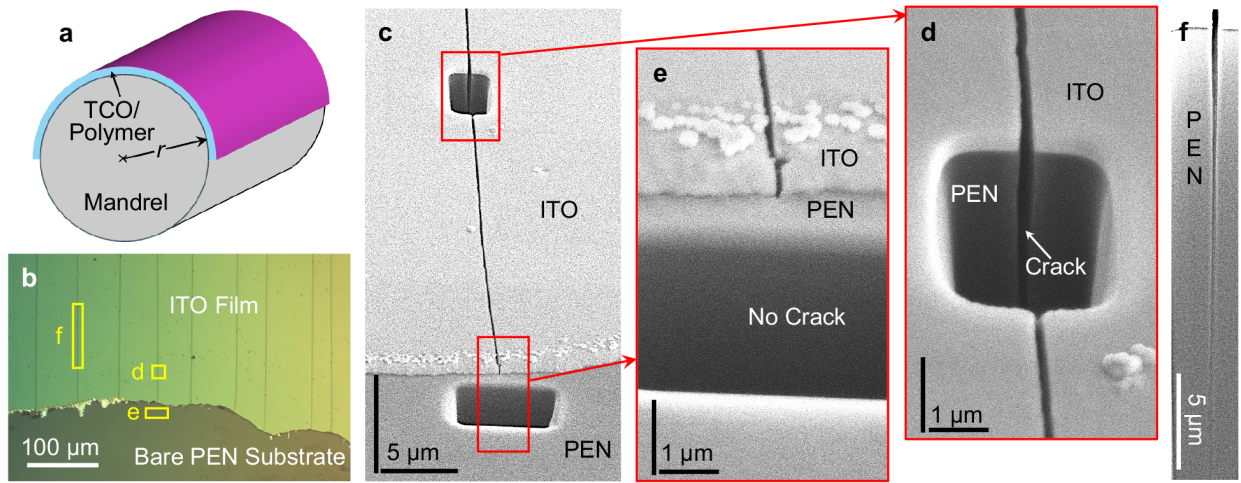

**Fig. 1. | Bending tests of ITO/PEN sheets. a**, Schematic illustration (not to scale) of the bending test of ITO/PEN sheets around a cylindrical mandrel of radius, $r$. **b**, Top-view optical microscope image of an ITO/PEN sheet bent to $r = 7$ mm ($\varepsilon_A = 0.0089$), in the bent state, showing 'channel' cracks in the ITO film and no cracks in the bare PEN substrate (bottom region). **c**, SEM image of the two FIB-cuts. (FIB-cutting was performed while in the bent state from regions similar to those indicated by the yellow rectangles in (**b**).) **d-e**, Corresponding higher-magnification SEM images of each of the FIB-cuts showing, in cross-sections, cracking of both ITO film and PEN substrate (**d**), and no cracking in the bare PEN substrate (**e**). **f**, Cross-sectional SEM image of a different FIB-cut showing deep cracking in the PEN substrate under ITO film. (All FIB-cut specimens were observed at a 52° forward tilt angle, hence vertical micron bars (representing depth) are longer than the horizontal ones, as indicated in (**c**)-(**f**).)

However, this assumption is upended by the scanning electron microscopy (SEM) observations of cross-sections created by focused ion beam (FIB) in Figs. 1c-d in the bent state, which surprisingly show cracking in the PEN substrate under the ITO film. It appears that the 'channel' cracks penetrate deep into the PEN substrate underneath. (No interfacial delamination was observed.) The cross-section SEM images in Figs. 1c-e confirm that the bare PEN substrate is uncracked. To further investigate the depth of the cracks in the PEN substrate, a longer 'trench' along the crack was FIB-cut, revealing deep cracks >25 μm (Fig. 1f), which is ~20% of the total substrate thickness. To confirm that the substrate cracking was not due to the FIB-cutting process itself, part of the ITO in a previously bent ($r$ = 7 mm) ITO/PEN sheet was removed by etching. Supplementary Figs. 2a,b present top-view SEM images of that specimen in the bent state, which clearly shows substrate cracking under the etched-away ITO. This type of pervasive, severe, and extensive substrate cracking is observed consistently in many 'channel' cracks within the same ITO/PEN specimen, and also in different ITO/PEN specimens. Similar substrate-cracking behavior is observed in ITO(~300 nm)/PET (~125 μm) (Fig. 3a) and ITO(~200 nm)/PI (~80 μm) (Fig. 3b) sheets, which are also commonly used in flexible electronic devices. Furthermore, substrate cracking is observed in multilayer devices built upon ITO/PEN sheets, *e.g.* flexible PSCs (see Supplementary Fig. 3), a promising new low-cost, light-weight photovoltaic PV technology.[6,27,28]. The key to the above insightful observations, which have eluded past studies, is that we observed the cross-sections, which were prepared by FIB-cutting and imaged in the SEM *in situ* in the bent state.

This revelation of substrate cracking in TCO/polymer sheets is very significant because it raises serious questions about the prevailing notion that high-toughness polymer substrates are immune to cracking. The current design, manufacturing, and operation of flexible electronic devices are based on this notion. Since the substrate serves as the foundational 'bedrock' of any flexible device, substrate cracking seriously undermines the mechanical integrity and reliability of the entire device, making it susceptible to cyclic fatigue and other time-dependent failure mechanisms, such as creep and environment-assisted cracking or degradation.

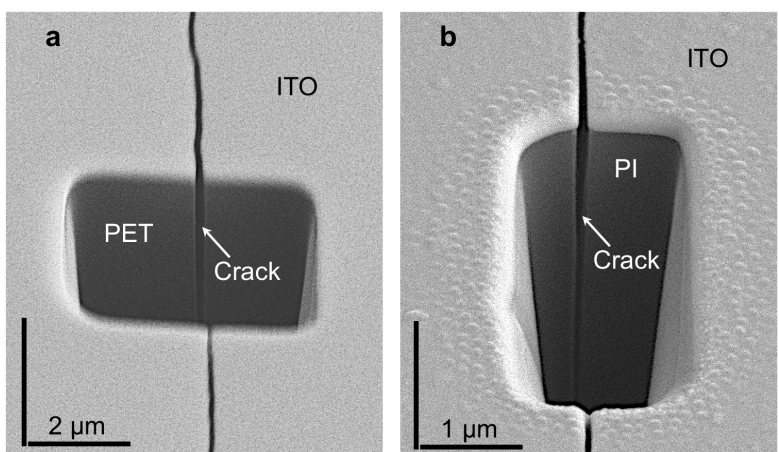

**Fig. 2. | Bending tests of ITO/PEN and ITO/PI sheets. a,b,** SEM images of FIB-cut cross-sections of 'channel' cracks in bent ITO/PET to $r$ = 7 mm ($\varepsilon_A$ = 0.0089) (**a**) and ITO/PI to $r$ = 5 mm ($\varepsilon_A$ = 0.0080) (**b**), in the bent state, showing cracking of both ITO films and substrates. (FIB-cutting was performed while in the bent state. All FIB-cut specimens were observed at a 52° forward tilt angle, hence vertical micron bar (representing depth) is longer than the horizontal one, as indicated.)

**Mechanics of substrate-cracking**

There have been only a few general studies reporting cracking in compliant, tough polymer substrates, such as PDMS and PET, underneath either stiff, ductile metal films (Au/PDMS,[29] Al/PDMS [30]) or stiff, brittle ceramic films (SiN$_x$/PET [31]). Cracking has also been reported in compliant, tough brass substrates underneath stiff, brittle TiN films.[32] However, in all those cases substrate cracking is modest, and to the best of our knowledge there have been no published reports of substrate cracking specifically in TCO/polymer sheets and pertinent flexible electronic devices.

To understand the genesis of the substrate-cracking phenomenon, we look back at some previous mechanics studies where the problem of cracking of compliant substrates underneath stiff films has been modeled.[33-36] For example, analytical modeling by Gecit [33] indicates that when a putative crack of depth, $c$, in a stiff film of thickness, $h$, (Al or steel) under uniaxial tension approaches the interface with a compliant substrate (epoxy or Al), the crack-driving stress intensity factor, $K_h$, increases dramatically relative to reference $K_o$ for no elastic mismatch between the film and the substrate. This is the result of concentrated shear stresses induced in the substrate at the crack-interface intersection due to the elastic mismatch,[37] and it is depicted schematically in Supplementary Fig. 4a. By corollary, when the elastic mismatch is reversed, *i.e.* compliant film (Al) on stiff substrate (steel), the $K_h$ diminishes.

Here it is instructive to define the elastic mismatch between the film (F) and the substrate (S) in terms of the Dundur's parameters, $\alpha$ and $\beta$, which incorporate not only their respective Young's moduli ($E_F$, $E_S$) but also shear moduli ($\mu_F$, $\mu_S$) and Poisson's ratios ($\nu_F$, $\nu_S$):[34]

$$\alpha = \frac{\bar{E_F} - \bar{E_S}}{\bar{E_F} + \bar{E_S}} \quad \text{and} \tag{1}$$

$$\beta = \frac{1}{2} \frac{\mu_F(1-2\nu_S) - \mu_S(1-2\nu_F)}{\mu_F(1-\nu_S) + \mu_S(1-\nu_F)}, \tag{2}$$

where $\bar{E} = E/(1-\nu^2)$ is the plane-strain Young's modulus. Using the Gecit model,[33] we calculate the normalized stress intensity factor $\hat{k}_n = K_h/K_o$ for a putative mode I tensile crack of normalized length $\hat{c} = c/h$ in ITO/PEN, ITO/PET, and ITO/PI film/substrate bilayer combinations, whose Dundur's parameters are listed in Supplementary Table 2 [32] and plotted in Fig. 3b (see Supplementary Note 2 for modeling details [33,38]). An incipient crack (small $\hat{c}$) at the surface of the film loaded in uniaxial tension (as in the bending test) will propagate when $K_h \geq K_{IC}$ condition is satisfied, where $K_{IC}$ is the fracture toughness of the film. In such a loading situation, $K_o \propto c^{0.5}$, and, thus, the crack is expected to propagate unstably.[39] The results in Fig. 3a show that, due to the presence of the compliant substrate, $K_h$ for a crack in the film increases with $\hat{c}$, beyond the conventional $c^{0.5}$-scaling increase represented by $\hat{k}_n = 1$. The $\hat{k}_n$ then rises dramatically as the crack approaches the interface ($\hat{c} \to 1$), and induces a stress intensity factor at the substrate surface that approaches infinity. The singularity implies that, in principle, the crack will continue to propagate into the substrate regardless of the substrate toughness. Figures 3d-e schematically depict this substrate-cracking behavior. By corollary, in the absence of the film, the substrate would otherwise not crack due its very high fracture toughness (Fig. 3c). Once the crack enters the substrate ($\hat{c} > 1$), $\hat{k}_n$ is expected to decrease as the crack propagates deeper into the substrate, an effect modeled by Thouless, *et al.*[36] In the case of bending, the applied $\sigma_A$ experienced by the crack tip also decreases as it propagates deeper into the substrate, and the stress goes to zero at the neutral axis, before becoming compressive.[13] Thus, cracks in the substrate are limited to ~20% of its thickness, but are sufficiently deep to cause cracking-induced degradation of the device.

**Substrate-cracking mitigation**

Using the Gecit model,[33] we have constructed a design map in Fig. 3b, where for $\alpha > -\beta$,

$\hat{k}_n$ is expected to increase dramatically as the crack approaches the interface ($\hat{c} \rightarrow 1$), resulting in substrate cracking. By corollary, $\hat{k}_n$ will decrease with $\hat{c} \rightarrow 1$ for $\alpha < -\beta$, resulting in crack arrest at the interface. As expected, ITO/PEN, ITO/PET, and ITO/PI combinations lie in the substrate-cracking region on the map in Fig. 3b. (IZO included in this map is another popular TCO that is used in this study, and its elastic properties are very similar to those of ITO (see Supplementary Table 1).) Since the much stiffer ITO (or IZO) film is key to the substrate-cracking problem, one can envision replacing the TCO with alternate, less stiff transparent-conducting electrodes made of polymers, metal meshes, or carbon-based nanomaterials.[40] However, those alternate electrodes are typically inferior to TCO in terms of the combination of high optical transparency and low sheet resistance they can offer.[40]

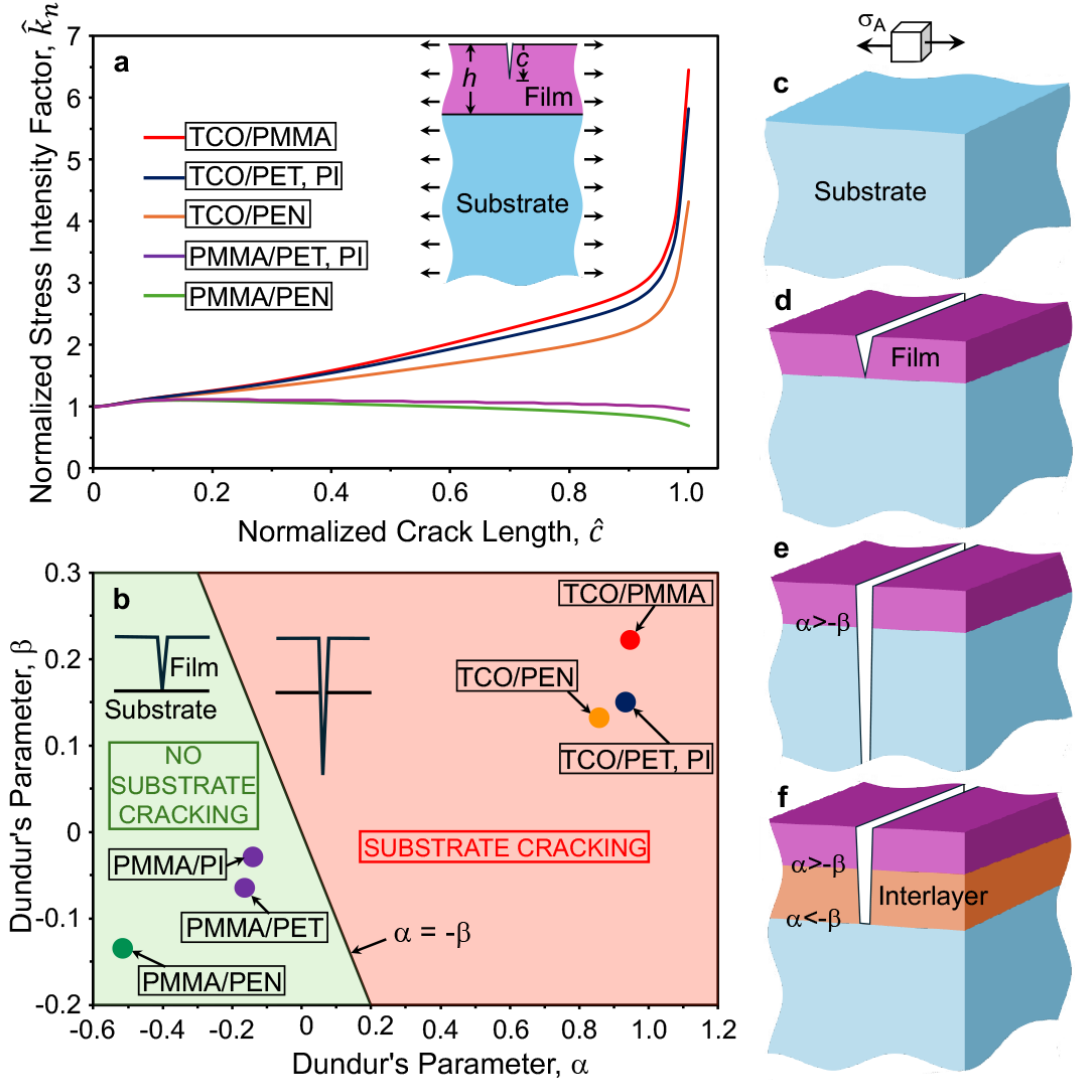

**Figure 3. | Modeling and mechanisms. a,** Plot of $\hat{k}_n$ ($= K_h/K_o$) as function of $\hat{c}$ ($= c/h$) for different film/substrate combinations relevant to this study. Inset: schematic illustration (not to scale) of the crack of depth, $c$, in the film of thickness, $h$. **b,** Design map of the two Dundur's parameters ($\alpha$ and $\beta$) for film/substrate showing the boundary ($\alpha = -\beta$) between substrate-cracking ($\alpha > -\beta$) and crack-arrest ($\alpha < -\beta$) regions. The film/substrate combinations in (**a**) are indicated on the map in (**b**). **c-f,** Schematic illustrations (not to scale) of cracking behavior under uniaxial tensile loading depicting no substrate cracking (**c**), 'channel' crack in film (**d**), extension of crack into the substrate (**e**), and substrate-cracking mitigation approach using an interlayer (**f**).

Using the design map in Fig. 3b, which provides general guidelines, we have designed a simple approach here for mitigating the substrate-cracking problem in generic TCO/polymer sheets for flexible electronic devices, without the need to eliminate the TCO. Our approach entails the use of a transparent, compliant interlayer between the TCO and polymer substrate, such that $\alpha < -\beta$ for the interlayer/substrate combination. To demonstrate the proof-of-concept, a polymethyl methacrylate (PMMA) interlayer ($E \sim 3$ GPa [19]) is chosen for insertion between IZO film ($E \sim 130$ GPa [22]) and PEN substrate ($E \sim 9$ GPa [15]) (Supplementary Table 1). Figure 3a plots $\hat{k}_n$ as a function of $\hat{c}$ for the IZO/PMMA and PMMA/PEN combinations, whose Dundur's parameters are listed in Supplementary Table 2 and plotted in Fig. 3b. For the IZO/PMMA case, a very rapid rise in $\hat{k}_n$ is observed as the crack approaches the interface, commensurate with the large $\alpha$ and $\beta$ for that combination. Thus, a putative crack in IZO film is expected to propagate into the PMMA interlayer underneath, as depicted schematically in Fig. 3f. However, in the case of the PMMA/PEN combination, for the same putative crack in the PMMA layer, $\hat{k}_n$ diminishes as the crack approaches the interface. This is due to the reversal of the concentrated shear stresses induced in the substrate due to the elastic mismatch (Supplementary Fig. 4b). Therefore, the crack is expected to be arrested, precluding any substrate cracking (Fig. 3f). While the approach of introducing interlayers between dissimilar materials for mitigating cracking has been discussed before (see *e.g.* [41,42]), this approach has eluded application to TCO/polymer sheets and pertinent flexible electronic devices.

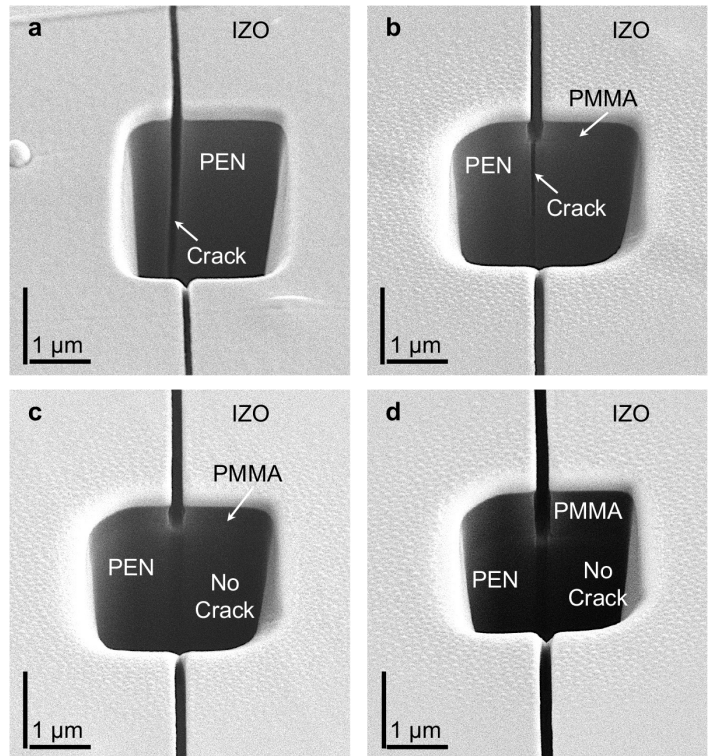

**Fig. 4. | Mitigation of substrate cracking. a-d**, SEM images of FIB-cut cross-sections of 'channel' cracks in combinations: IZO/PEN (**a**), IZO/PMMA(250 nm)/PEN (**b**), IZO/PMMA(300 nm)/PEN (**c**), and IZO/PMMA(600 nm)/PEN (**d**), bent to $r = 7$ mm ($\varepsilon_A = 0.0089$), in the bent state. A thin $SiO_2$ layer ($\sim 20$ nm) was used on top of PMMA. (FIB cuts were made while in the bent state. All FIB-cut specimens observed at a 52° forward tilt angle, hence vertical micron bars (representing depth) are longer than the horizontal ones, as indicated in (**a**)-(**d**).)

To test this hypothesis experimentally, PMMA layers of different thicknesses (~250, ~300, ~400, and ~600 nm) were spin-coated onto PEN substrates. Unfortunately, PMMA is unable to withstand the processing conditions of the subsequent IZO deposition. Therefore, an ultrathin, transparent layer of amorphous $SiO_2$ (~20 nm) is thermally evaporated on the PMMA surface to protect it. Considering the thinness of the $SiO_2$, it is not expected to interfere with this mitigation approach. IZO (~250 nm thickness) was then sputter-deposited on top of the $SiO_2$. In the case of no-PMMA, substrate cracking is observed in the FIB-cut cross-section in Fig. 4a, which is similar to that in the ITO/PEN case (Fig. 1d). Similarly, in the case of 250-nm PMMA interlayer, cracking of the PEN substrate is observed in Fig. 4b. However, dramatic arrest of 'channel' cracks at the PMMA/PEN interface is observed in Fig. 4c with a ~300 nm PMMA interlayer. Similar crack-arrest behavior is observed in Fig. 4d with a thicker PMMA interlayer (~600 nm). In related experiments, instead of the $SiO_2$ protective layer, an Au layer (~10 nm) was deposited, which allowed better contrast delineation of the PMMA layer in the SEM; the same type of crack-arrest behavior is observed with a ~400 nm PMMA interlayer (Supplementary Fig. 5). This indicates that a relatively thin interlayer is sufficient for suppressing substrate cracking. Thus, the design map in Fig. 3b can be used as a general guide for choosing a suitable transparent interlayer between the TCO film and the substrate to prevent substrate cracking in those devices. Of course, the transparency requirement on the interlayer is relaxed for use in otherwise opaque flexible electronic devices.

**Electrical properties**

The relative change in the resistance ($\Delta R/R_O$) of the ITO film on PEN substrate was measured at different bending radii, $r$, in the bent state (Supplementary Fig. 6a), where the resistance $R$ is in the direction perpendicular to the bending axis (Supplementary Fig. 6a inset), $R_O$ is the initial resistance in the flat state, and $\Delta R = R\text{-}R_O$. Beyond a critical bending radius ($r = 10$ mm), ITO's $\Delta R/R_O$ increases monotonically with decreasing $r$. Supplementary Fig. 6b plots the corresponding measured density of 'channel' cracks as a function of decreasing $r$, in the bent state, which also increases monotonically. It is assumed that the increase in the crack density in the ITO film is responsible for the increase in the relative change in the resistance in the bent state. However, when the ITO/PEN sheet is returned to the flat state (*i.e.* one bending cycle), $\Delta R/R_O$ typically approaches zero. This is attributed to the fact that the mating walls of 'channel' cracks in the ITO can make good contact in the flat state, although the cracks are not healed.

In light of the substrate-cracking discovery reported here, it is hypothesized that when the TCO/polymer sheet is subjected to cyclic bending (*i.e.* multiple cycles), the propensity for the substrate cracks to close properly when returned to the flat state will decrease progressively. This is most likely due to the accumulation of debris within the substrate cracks, their propagation deeper into the substrate, and growing misalignment of the mating crack walls, with increasing number of cycles. This, in turn, is likely to progressively prevent the TCO crack walls from making good contact, thereby resulting in increasing $\Delta R/R_O$ (flat state) with number of cycles, $n$. It is also hypothesized that in the case where substrate cracking is eliminated, *e.g.* in IZO/PMMA(300 nm)/PEN, the flat-state $\Delta R/R_O$ may not increase with increasing $n$. These hypotheses are illustrated schematically in Supplementary Fig. 7. To verify these hypotheses, $\Delta R/R_O$ of IZO/PEN sheets was measured in the flat state (Fig. 5) after subjecting them to bending cycles in the range $n = 1$ to $10^4$, at bending radius $r = 7$ mm ($\varepsilon_A = 0.0089$) using a testing method described elsewhere.[43] Figure 5 plots these data, which clearly show progressive increase in flat-state $\Delta R/R_O$ with number of bending cycles, $n$, which reaches up to ~1,200 after $10^4$ bending cycles. In contrast, no appreciable change in $\Delta R/R_O$ could be measured in IZO/PMMA(300 nm)/PEN in the flat state, even after $10^4$ cycles where $\Delta R / R_O \sim 0.5$. These results reinforce the critical importance of substrate cracking in the degradation of TCO electrical properties. Since electrical continuity of the TCO film is key to the proper functioning of any multilayer device built upon TCO/polymer sheets, substrate cracking is highly detrimental for the durability and reliability of such devices.

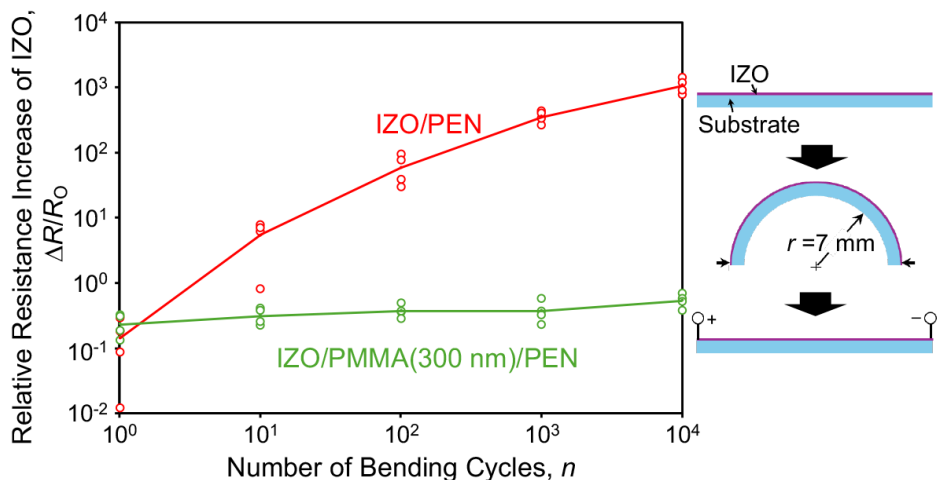

**Fig. 5 | Cyclic bending and electrical properties.** The relative increase in the DC electrical resistance of IZO film on PEN substrate, where the substrate cracks in bending ($r = 7$ mm; $\varepsilon_A = 0.0089$), in the flat state as a function of bending cycles, $n$. Corresponding measurements of IZO on PMMA(300 nm)/PEN substrate, where the substrate does not crack in bending. Measurements from four different specimens for each material system reported. The solid lines connect the averages. The schematic illustration (not to scale) on the right depicts one bending cycle: flat→bent→flat.

In closing, while the severity and extent of polymer-substrate cracking revealed here will depend on the materials, their relative thicknesses, and applied tensile stresses, the generic substrate-cracking mitigation approach of introducing a more compliant interlayer above the substrate is potentially applicable to myriad flexible electronic devices, and also other multilayer systems in general. It is imperative that consideration of polymer substrate cracking, and its mitigation, become integral part of future design, manufacturing, and operation of flexible electronic devices with improved durability and reliability.

## Methods

Materials

All reagents/materials were used as-received commercially without further purification, which include: lead(II) iodide ($PbI_2$; 99%, Sigma Aldrich, USA), formamidinium iodide (FAI; >99.99%, Greatcell Solar Materials, Australia), lead(II) bromide ($PbBr_2$; Materials, Australia), cesium iodide (CsI; 99.999%, Sigma Aldrich, USA), methylammonium bromide (MABr; >99.99%, Greatcell Solar Materials, Australia), methylammonium chloride (MACl; >99.99%, Greatcell Solar Materials, Australia), N,N-dimethylformamide (DMF; 99.8%, anhydrous, Sigma Aldrich, USA), dimethyl sulfoxide (DMSO; ≥99.9%, anhydrous, Sigma Aldrich, USA), ethanol (>99.5%; Sigma Aldrich, USA), acetone (>99.5%, VWR, USA), $SnO_2$ colloidal dispersion ($SnO_2$-CD; 15 wt % in $H_2O$, Alfa Aesar, USA), isopropanol (IPA; 99.5%, anhydrous, Sigma Aldrich, USA), bis(trifluoromethane)sulfonimide lithium salt (Li-TFSI; 99.95%, Sigma Aldrich, USA), 4-tert-butylpyridine (tBP; 98%, Sigma Aldrich, USA), acetonitrile (AcN; 99.8%, anhydrous, Sigma Aldrich, USA), chlorobenzene (CB; 99.8%, anhydrous, Sigma Aldrich, USA), spiro-OMeTAD (>99.8%, Lumtec, Taiwan), polymethylmethacrylate solution (PMMA; Sigma Aldrich (182265), USA), polyethylene naphthalate sheets (PEN; Goodfellow Corp., USA), polyimide sheets (PI; Kapton, Kolon Industried, S. Korea), indium-tin-oxide-coated polyethylene terephthalate substrates (ITO/PET; Sigma Aldrich, USA), and ITO-coated PEN substrates (ITO/PEN; Peccell, Japan).

### Deposition of PMMA Thin Films

Solution of PMMA in CB (5 wt%, heated at 70 °C overnight) was spin-coated on PEN substrates under ambient conditions at 6,000, 4,000, 2,000, or 1,000 rpm, each with a ramp of 500 rpm·s$^{-1}$, for 30 s to achieve a thicknesses of ~250, ~300, ~400, or ~600 nm, respectively.

### Deposition of SiO₂ and Au Thin Films

SiO$_2$ thin films of ~20 nm thickness were deposited on top of the PMMA thin films at room temperature using an electron-beam evaporator (Ångstrom, USA), at a rate of 1 Å·s$^{-1}$. In related experiments, instead of SiO$_2$ thin films, Au thin films (~10 nm) were deposited by thermal evaporation (Ångstrom, USA).

### Deposition of ITO Thin Films

ITO thin films (~200 nm thickness) were deposited on PI substrates (~80 μm thickness) using a radio frequency (RF) linear sputtering system (Kenosistec, Italy). The chamber was first evacuated to $5\times10^{-6}$ Torr, followed by a pre-sputtering phase to cleanse the target and enhance the process reproducibility. Deposition occurred under $1.1\times10^{-3}$ Torr pressure with an Ar flow of 40 SCCM and a power density of 0.39 W·cm$^{-2}$. During the deposition, the substrate was moved laterally to ensure uniform deposition.

### Deposition of IZO Thin Films

IZO thin films (~250 nm thickness) were deposited on the SiO$_2$/PMMA/PEN or Au/PMMA/PEN sheets using DC-magnetron sputtering at room temperature using a IZO ceramic target (90 wt.% In$_2$O$_3$-10 wt.% ZnO). Specimens were attached to a metal mask and placed in the chamber, which was then evacuated until the pressure dropped below $5\times10^{-6}$ Torr. Ar gas was passed through the gas mass flow controller into the chamber at 8 SCCM. The power supply for the plasma generator was then turned on and the working pressure was varied such that the DC bias was 280-300 V. After comparing different deposition times and measuring the respective film thickness, the deposition rate was determined to be ~0.25 Å·s$^{-1}$. Etching of IZO films, as needed, was performed using the Zn-HCl chemical-etching method.

### Flexible PSCs Fabrication

The ITO/PEN sheets were patterned using the Zn-HCl chemical-etching method. They were then sonicated in a solution of IPA, acetone, and DI (1:1:1 v/v) for 20 min. The substrates were subsequently dried under flowing dry N$_2$ and attached to an appropriately-sized microscope glass slide using tape. The substrates were treated under ultraviolet ozone (UVO) for 30-35 min and transferred to a controlled-humidity (<10% RH) glovebox. Immediately after this, an SnO$_2$ solution (SnO$_2$-CD in DI water; 1:3 v/v, filtered with a 0.22 mm PTFE filter) was spin-coated at 4,000 rpm with a ramp of 1,000 rpm·s$^{-1}$ for 30 s, followed by drying at 120 °C for 60 min on a hotplate.

The metal-halide perovskite (MHP) precursor solution (1.61 M) was prepared by dissolving 742.2 mg of PbI$_2$, 224.4 mg of FAI, 16.2 mg of MABr, 20.3 mg of MACl, and 19.8 mg of CsI in a mixture of 800 μL DMF and 200 μL DMSO. This is to result in thin films of final nominal composition of Cs$_{0.05}$MA$_{0.13}$FA$_{0.82}$Pb(I$_{2.9}$Br$_{0.1}$). This precursor solution was stirred for 2 h at room temperature and filtered before use. After treating the substrates under UVO for 15 min, the MHP layer was spin-coated on the prepared substrate at 1,000 rpm, with a ramp of 200 rpm·s$^{-1}$ for 10 s followed by 5,000 rpm, with a ramp of 1,000 rpm·s$^{-1}$, for 30 s. Anhydrous CB (0.2 mL) was dripped in a single stream onto the rotating substrate 15 s before the end. Subsequently, the film was annealed for 30-40 min at 100 °C under ~17% RH on a hotplate. The humidity was then reduced to <10% RH. The Spiro-OMeTAD layer was then spin-coated at 4,000 rpm, with a ramp of 2,000 rpm·s$^{-1}$, for 20 s using a solution of 90 mg Spiro-OMeTAD, 39.5 μL of tBP, 23 μL of Li-TFSI (520 mg dissolved in 1 mL of acetonitrile) in 1 mL CB (filtered with 0.22 μm PTFE filter). All processing was performed in a humidity-controlled ambient-air glovebox. Finally, 70-nm thick Au electrode was deposited by thermal evaporation. The flexible PSCs were then detached from the glass slide.

### Bending Tests and FIB Cutting

Focused ion beam (FIB) cutting of cross-sections was performed using a Helios 600 instrument

(ThermoFisher, USA) with all the specimens being in the bent state. The different flexible specimens were carefully wrapped around custom-made acrylic cylindrical mandrels with $r$ ranging from 5 to 10 mm. In each case, the length of the substrate was $\sim\pi r$. After sputtering a thin layer of conductive carbon, the mandrels were carefully attached to the FIB stage using carbon tape. Additional copper tapes were used to hold the sample down, as well as to reduce any charging effects. A beam current of 0.46-0.92 nA was used to make the FIB cuts, in cleaning cross-section mode, without any Pt deposition. It took approximately 3-5 minutes to make each FIB cut.

<u>Characterization</u>

All scanning electron microscope (SEM; Quattro S, ThermoFisher, USA) observations, top view and cross-sectional view, were performed while the specimens were in the bent state. The same procedure as above was used to mount the bent specimens onto the SEM stage.

An optical microscope (Eclipse LV100, Nikon, USA) was used to make top-view observations, and also to measure the crack densities while the specimens were in the bent state. The same procedure as above was used to mount the bent specimens onto the optical microscope stage. The number of cracks in an image was divided by the corrected image length, to account for the curvature, and this number is referred to as the crack density. For each $r$ condition, a minimum of three specimens were used for this measurement.

<u>Electrical Properties Measurements</u>

A multimeter (179, Fluke, USA) was used to measure the DC electrical resistance of the ITO film in ITO/PEN. Copper tapes were attached to the ends of the ITO film, and the initial resistance ($R_O$) in the flat state was carefully measured, which was found to be very close to a value calculated from its specified sheet resistance of the ITO film (13-15 $\Omega/$ ). The ITO/PEN specimens were manually bent on the acrylic cylindrical mandrels of different $r$ (5, 7, or 10 mm). Copper tapes were attached to the flat part of the ITO film at either ends, and they were connected to the multimeter to measure the resistance ($R$) of the ITO film in the bent state. For each $r$ condition, about four specimens were used for this measurement. All bending and measurement procedures were performed in a dark, dry-air glovebox (<5% RH).

Cyclic bending tests were performed on IZO/PEN and IZO/PMMA(300 nm)/PEN sheets in air (~25 °C, ~25% RH) using an automatic cyclic mechanical tester (PR-BDM-100, Puri, China), where the acrylic cylindrical mandrel was of radius $r$=7 mm; each bending cycle is: flat→bent→flat.[43] Testing was performed for 10,000 cycles, at the rate of ~1 cycle·s$^{-1}$. Testing was interrupted periodically, and the DC electrical resistance of the IZO films in the flat state were measured using the method described above. The initial resistance ($R_O$) values of the IZO films in the flat state were measured before the commencement of cyclic bending tests.

# Data availability

The data that support the findings of this study are available from the corresponding author on reasonable request.

# References

1.  Corzo, D., Tostado-Blázquez, G. & Baran, D. Flexible electronics: Status, challenges and opportunities. *Front. Electron.* **1**, 594003 (2020).
2.  Wang, Y., Xu, C., Yu, X., Zhang, H. & Han, M. Multilayer flexible electronics: Manufacturing approaches and applications. *Mater. Today Phys.* **23**, 100647 (2022).
3.  *https://www.snsinsider.com/reports/flexible-electronics-market-4778* (2024).
4.  Subudhi, P. & Punetha, D. Progress, challenges, and perspectives on polymer substrates for emerging flexible solar cells: A holistic panoramic review. *Prog. Photovolt.* **31**, 753-789 (2024).

5.   Yan, Y., Duan, B., Ru, M., Gu, Q., Li, S. & Zhao, W. Toward flexible and stretchable organic solar cells: A comprehensive review of transparent conductive electrodes, photoactive materials, and device performance. *Adv. Energy Mater.* **15**, 2404233 (2025).

6.   Padture, N. P. The promise of metal-halide-perovskite solar photovoltaics: A brief review. *MRS Bull.* **48**, 983-998 (2023).

7.   Koden, M. *OLED Displays and Lighting.* (John Wiley & Sons, 2017).

8.   Ouyang, C., Liu, D., He, K. & Kang, J. Recent advances in touch sensors for flexible displays. *IEEE J. Nanotechnol.* **4**, 36-46 (2023).

9.   Fang, H. J., Zhao, Z. T., Wu, W. T. & Wang, H. Progress in flexible electrochromic devices. *J. Inorg. Mater.* **36**, 140-151 (2021).

10.  Morales-Masis, M., DeWolf, S., Woods-Robinson, R., Ager, J. W. & Ballif, C. Transparent electrodes for efficient optoelectronics. *Adv. Electron. Mater.* **3**, 1600529 (2017).

11.  Zhao, Z., Fu, H., Tang, R., Zhang, B., Chen, Y. & Jiang, J. Failure mechanisms in flexible electronics. *Intl. J. Smart Nano Mater.* **14**, 510-565 (2023).

12.  Zhu, Y. & Lu, N. *Mechanics of Flexible and Stretchable Electronics.* (Wiley-VCH, 2025).

13.  Fukuda, K., Sun, L., Du, B., Takakuwa, M., Wang, J., Someya, T., Marsal, L. F., Zhou, Y., Chen, Y., Chen, H., Silva, S. R. P., Baran, D., Castriotta, L. A., Brown, T. M., Yang, C., Li, W., Ho-Baillie, A. W. Y., Österberg, T., Padture, N. P., Forberich, K., Brabec, C. J. & Almora, O. A bending test protocol for characterizing the mechanical performance of flexible photovoltaics. *Nature Energy* **9**, 1335-1343 (2024).

14.  Bárány, T., Czigány, T. & Karger-Kocsis, J. Application of the essential work of fracture (EWF) concept for polymers, related blends and composites: A review. *Prog. Polymer Sci.* **35**, 1257-1287 (2010).

15.  Arkhireyeva, A. & Hashemi, S. Fracture behavior of polyethylene napthalate (PEN). *Polymers* **43**, 289-300 (2002).

16.  Jung, H. S., Eun, K., Kim, Y. T., Lee, E. K. & Choa, S.-H. Experimental and numerical investigation of flexibility of ITO electrode for application in flexible electronic devices. *Microsys. Technol.* **23**, 1961-1970 (2017).

17.  Dai, Z. & Padture, N. P. Challenges and opportunities for the mechanical reliability of metal-halide perovskites and photovoltaics. *Nature Energy* **8**, 1319-1327 (2023).

18.  Zhou, J., Zhang, X., Zhang, X., Zhang, W., Li, J., Chen, Y., Liu, H. & Yan, Y. Mechanical properties of tensile cracking in indium tin oxide films on polycarbonate substrates. *Coatings* **12**, 538 (2022).

19.  Ishiyama, C. & Higo, Y. Effects of humidity on young's modulus in poly(methyl methacrylate). *Journal of Polymer Science: Part B: Polymer Physics* **40**, 460-465 (2002).

20.  Kono, R. Dynamic bulk viscosity of polystyrene and polymethyl methacrylate. *J. Phys. Soc. Jpn.* **15**, 718-725 (1960).

21.  Rittel, D. & Maigre, H. An investigation of dynamic crack initiation in PMMA. *Mech. Mater.* **23**, 229-239 (1996).

22.  Zhou, J., Zhang, X., Zhang, X., Zhang, W., Chen, Y., Shi, H. & Yan, Y. Effects of dielectric layer on ductility for dielectric/au/dielectric multilayers on polycarbonate substrate. *J. Phys. D: Appl. Phys.* **56**, 435302 (2023).

23.  Lu, H. & Menary, G. Determination of young's modulus of PET sheets from Lamb wave velocity measurement. *Exptl. Mech.* **64**, 377-391 (2024).

24.  Chan, W. Y. F. & Williams, J. G. Determination of the fracture toughness of polymeric films by the essential work method. *Polymer* **35**, 1666-1672 (1994).

25. He, W., Goudeau, P., Bourhis, E. L., Renault, P.-O., Dupré, J. C., Doumalin, P. & Wang, S. Study on Young's modulus of thin films on Kapton by microtensile testing combined with dual dic system. *Surf. Coat. Technol.* **308**, 273-279 (2016).

26. Bauer, C. L. & Farris, R. J. Determination of Poisson's ratio for polyimide films. *Polymer Engr. Sci.* **29**, 1107-1110 (1989).

27. Gao, Y., Huang, K., Long, C., Ding, Y., Chang, J., Zhang, D., Etgar, L., Liu, M., Zhang, J. & Yang, J. Flexible perovskite solar cells: From materials and device architectures to applications. *ACS Energy Lett.* **7**, 1412-1445 (2022).

28. Dai, Z., Li, S., Liu, X., Chen, M., Athanasiou, C. E., Sheldon, B. W., Gao, H., Guo, P. & Padture, N. P. Dual-interface reinforced flexible perovskite solar cells for enhanced performance and mechanical reliability. *Adv. Mater.* **34**, 2205301 (2022).

29. Douville, N. J., Liu, Z., Takayama, S. & Thouless, M. D. Fracture of metal coated elastomers. *Soft Matter* **7**, 6493-6500 (2011).

30. Badrudin, S. I., Noor, M. M., Samad, M. I. A., Zakaria, N. S. N., Yunas, J. & Latif, R. Eliminating surface cracks in metal film-polymer substrate for reliable flexible piezoelectric devices. *Intl. J. Engr. Sci. Technol.* **50**, 101617 (2024).

31. Kim, K., Luo, H., Zhu, T., Pierron, O. N. & Graham, S. Influence of polymer substrate damage on the time dependent cracking of SiN$_x$ barrier films. *Sci. Rep.* **8**, 4560 (2018).

32. Gao, T., Qiao, L., Pang, X. & Volinsky, A. A. Brittle film-induced cracking of ductile substrates. *Acta Mater.* **99**, 273-280 (2015).

33. Gecit, M. R. Fracture of a surface layer bonded to a half space. *Intl. J. Engr. Sci.* **17**, 287-295 (1979).

34. Hutchinson, J. W. & Suo, Z. Mixed-mode cracking in layered structures. *Advances in Appl. Mech.* **29**, 63-191 (1991).

35. Beuth, J. L. Cracking of thin bonded films in residual tension. *Intl. J. Solids Struct.* **28**, 1657-1675 (1992).

36. Thouless, M. D., Li, Z., Douville, N. J. & Takayama, S. Periodic cracking of films supported on compliant substrates. *J. Mech. Phys. Solids* **59**, 1927-1937 (2011).

37. Beuth, J. L. & Klingbell, N. W. Cracking of thin films bonded to elastic-plastic substrates. *J. Mech. Phys. Solids* **44**, 1411-1428 (1996).

38. Bower, A. F. Applied Mechanics of Solids. *https://solidmechanics.org* (2025).

39. Lawn, B. R. *Fracture of Brittle Solids*. (Cambridge University Press, 1993).

40. Xu, Y., Lin, Z., Wei, W., Hao, Y., Liu, S., Ouyang, J. & Chang, J. Recent progress of electrode materials for flexible perovskite solar cells. *Nano-Micro Letters* **14**, 117 (2024).

41. Suresh, S., Sugimura, Y. & Ogawa, T. Fatigue cracking in materials with brittle surface coatings. *Scripta Mater.* **29**, 237-242 (1993).

42. Freund, L. B. & Suresh, S. *Thin Film Materials: Stress, Defect Formation and Surface Evolution*. (Cambridge University Press, 2003).

43. Dai, Z., Li, S., Liu, X., Chen, M., Athanasiou, C. E., Sheldon, B. W., Gao, H., Guo, P. & Padture, N. P. Dual-interface reinforced flexible perovskite solar cells for enhanced performance and mechanical reliability. *Adv. Mater.* **34**, 2205301 (2022).

## Acknowledgements

Experimental assistance from H.F. Garces and M. Poma is gratefully acknowledged. The work at Brown University was supported by the U.S. Department of Energy (DOE) Office of Energy Efficiency and Renewable Energy (EERE) under the Solar Energy Technology Office (SETO) (Award No. DE-EE0009511), DOE Basic Energy Sciences (BES) (Award No. DE-SC0025180), and the U.S. National Science Foundation (NSF) (Grant No. DMR-2102210). The views expressed in the article do not necessarily represent the views of the DOE or the U.S. Government. S.K. and H.K. gratefully acknowledge the support of the U.S. Office of Naval Research (ONR) (Award Nos. N00014-21-1-2851, N00014-24-1-2200, and N00014-21-1-2054). Additional support from ONR (Award Nos. N00014-21-1-2815 and N00014-23-1-2688) is gratefully acknowledged. S.K. is also grateful for the support she received through the James R. Rice Graduate Fellowship in Solid Mechanics and the Miss Abbott's School Alumnae Fellowship. S.S. acknowledges the support from Brown University as part of his Professor-at-Large appointment. The work at Yale University was supported by NSF (Grant No. CBET-2315077) and NSF-GRFP (Grant No. DGE-2139841). The work at University of Rome was supported by the JUMP INTO SPACE project, funded from the European Innovation Council (EIC) under grant agreement No 101162377. This publication reflects only the author's views and the European Union is not liable for any use that may be made of the information contained therein. F.B. thanks G. Koch and F. De Rossi for the support in preparing the PI specimens. E.M. and A.D.C. acknowledge the support of MASE (Ministero dell'Ambiente e della Sicurezza Energetica) in the framework of the Operating Agreement with ENEA for Research on the Electric System (RdS) 2025-2027.

# Author information
*Authors and Affiliations*

*School of Engineering, Brown University, Providence, RI 02912, USA*
Anush Ranka, Madhuja Layek, Sayaka Kochiyama, Cristina López-Pernia, Alicia M. Chandler, Subra Suresh, David C. Paine, Haneesh Kesari, Nitin P. Padture

*Department of Chemical and Environmental Engineering, Yale University, New Haven, CT 06520, USA*
Conrad A. Kocoj, Peijun Guo

*Energy Sciences Institute, Yale University, West Haven, CT 06516, USA*
Conrad A. Kocoj, Peijun Guo

*Centre for Hybrid and Organic Solar Energy, Department of Electronic Engineering, University of Rome Tor Vergata, Rome 00133, Italy*
Erica Magliano, Aldo Di Carlo, Francesca Brunetti

*Instituto di Struttura della Materia, National Research Council, Rome 00133, Italy*
Aldo DiCarlo

*Contributions*
N.P.P. conceived and supervised the research. N.P.P. and A.R. designed the research. A.R. performed bulk of the experimental work, including synthesis, processing, fabrication, characterization, measurements and testing, with contributions from M.L., C.L-P., A.C., C.K., E.M., A.D.C., F.B., P.G., and D.P. The modeling work and the interpretation of the results was performed by S.K. and H.K. The data were interpreted by N.P.P., A.R., H.K., and S.S. The manuscript was written by N.P.P. and A.R., with contributions from the other authors.

# Ethics declarations
*Competing Interests*
The authors declare no competing interests. A provisional application for a patent has been filed.

# **Supplementary Information**

## **Supplementary Note 1: Strain in Multilayer Bent Device**

In a multilayer device with $n$ layers bent to radius $r$, the strain is given by $\varepsilon_{ML}=(z-b)/r$, where $z$ is the position along the thickness, and $b$ is the neutral mechanical plane (zero-strain position), given by:[13]

$$b = \frac{\sum_{i=1}^{n} E_i t_i \left[\sum_{j=1}^{i} t_j - \frac{t_i}{2}\right]}{\sum_{i=1}^{n} E_i t_i}, \text{(Eqn. S1)}$$

where $E_i$ and $t_i$ denote the Young's moduli and thicknesses, respectively, of the individual $i^{\text{th}}$ layer.

## **Supplementary Note 2: Fracture Mechanics Modeling**

The plots in Fig. 3a were generated using the model by Gecit [33] with some necessary minor corrections and modifications to the original analysis. Here the normalized stress intensity factor is defined as:

$$\hat{k}_n = \frac{K_h}{K_o}, \text{(Eqn. S2)}$$

where $K_h$ is the stress intensity factor at the tip of a putative crack of depth $c$ in the film of thickness $h$ (see Fig. 3a inset). $K_o$ is the reference stress intensity factor for the case where there is no elastic mismatch between the film and the substrate. The normalized crack length, $\hat{c}$, is defined as $c/h$ such that $0 \leq \hat{c} < 1$.

The plots in Fig. 3a are given as the graphs of the function $\hat{k}_n[\cdot]: [0,1) \rightarrow \mathbb{R}$,

$$\hat{k}_n[\hat{c}] \coloneqq \theta[\hat{c}; 1], \text{(Eqn. S3)}$$

where $\theta[\hat{c}; \cdot]: (0,1) \rightarrow \mathbb{R}$, is obtained by interpolating the solution to the system of linear equations:

$$\sum_{i=1}^{n} W_i\, \theta[\hat{c}; \tau_i] \left[\frac{1}{\tau_i - \omega_j} + \hat{c}\,\hat{k}[\hat{c}\omega_j, \hat{c}\tau_i]\right] = -\pi \qquad j = 1, \dots, n, \text{(Eqn. S4)}$$

provided,

$$\tau_i = \cos\left[\frac{(2i-1)\pi}{4n+2}\right] \qquad i = 1, \dots, 2n+1, \text{(Eqn. S5)}$$

$$W_i = \frac{\pi}{2n+1} \qquad i = 1, \dots, 2n+1, \text{(Eqn. S6)}$$

$$\omega_j = \cos\left[\frac{j\pi}{2n+1}\right] \qquad j = 1, \dots, 2n+1, \text{(Eqn. S7)}$$

and $\hat{k}[\cdot,\cdot]: (0,\hat{c}) \times (0,\hat{c}) \rightarrow \mathbb{R}$ given by:

$$\hat{k}[\hat{x}_F, \hat{t}] \coloneqq \frac{1}{\hat{t} + \hat{x}_F} + \int_0^{\infty} \widehat{\Pi}[\hat{x}_F, \hat{t}, z]\ dz. \text{ (Eqn. S8)}$$

The function $\widehat{\Pi}[\cdot,\cdot,\cdot]$ appearing in the above definition is given below. Obtaining the interpolated expression for function $\theta[\hat{c}; \cdot]$ as outlined above, the right-hand side of Eqn. S3 can be evaluated by taking the limit of $\theta[\hat{c}; \tau]$ as $\tau \rightarrow 1$.

In Eqn. S8, $\widehat{\Pi}: (0,\hat{c}) \times (0,\hat{c}) \times \mathbb{R}_{\geq 0} \rightarrow \mathbb{R}$:

$$\widehat{\Pi}[\hat{x}_F, \hat{t}, z] := \sum_{j=1}^{5} \left[ \left( \delta_{ij} + 2a_j[z] - z\hat{x}_F b_j[z] \right) \cosh(z\hat{x}_F) \right.$$
$$\left. + \left( b_j[z] + z\hat{x}_F a_j[z] \right) \sinh(z\hat{x}_F) \right] M_j[z, \hat{t}] , \text{(Eqn. S9)}$$

where $\delta_{ij}$ is the Kronecker delta,[38] and:

$a_1[z] := (-\lambda\rho + 2\lambda\gamma e^{-2z} + \gamma\eta e^{-4z})/d[z]$, (Eqn. S10)
$a_2[z] := \xi(2\lambda z - \epsilon - ((2\gamma z - 1)\kappa_S + \kappa_F + \kappa_S)e^{-2z})e^{-z}/d[z]$, (Eqn. S11)
$a_3[z] := \xi(2\lambda(z-1) + \epsilon + (2\gamma(z+1)\kappa_S - \kappa_F + \kappa_S)e^{-2z})e^{-z}/d[z]$, (Eqn. S12)
$a_4[z] := -(\lambda(2z-1) - \rho + (\gamma(2z+1) - \eta)e^{-2z})e^{-z}/d[z]$, (Eqn. S13)
$a_5[z] = (-\lambda(2z-1) - \rho + (\gamma(2z+1) + \eta)e^{-2z})e^{-z}/d[z]$. (Eqn. S14)

$b_1[z] := (\lambda\rho + 4\lambda\gamma z e^{-2z} + \gamma\eta e^{-4z})/d[z]$, (Eqn. S15)
$b_2[z] := -\xi(2\lambda(z+1) - \epsilon + (2\gamma(z-1)\kappa_S + \kappa_F - \kappa_S)e^{-2z})e^{-z}/d[z]$, (Eqn. S16)
$b_3[z] := \xi(-2\lambda z - \epsilon + ((2\gamma z + 1)\kappa_S - \kappa_F)e^{-2z})e^{-z}/d[z]$, (Eqn. S17)
$b_4[z] := (\lambda(2z+1) - \rho + (\gamma(-2z+1) + \eta)e^{-2z})e^{-z}/d[z]$, (Eqn. S18)
$b_5[z] := (\lambda(2z+1) + \rho + (\gamma(2z-1) + \eta)e^{-2z})e^{-z}/d[z]$. (Eqn. S19)

$d[z] := -\lambda\rho + (\lambda\gamma(4z^2+1) - \rho\eta)e^{-2z} + \gamma\eta e^{-4z}$. (Eqn. S20)

Here: $\xi = \mu_F/\mu_S$, $\lambda = 1 + \xi\kappa_S$, $\rho = \xi + \kappa_F, \gamma = \xi - 1$, $\eta = \kappa_F - \xi\kappa_S$, and $\epsilon = 1 - \kappa_F\kappa_S$, where $\kappa_F = 3 - 4\nu_F$ and $\kappa_F = (3 - \nu_F)/(1 + \nu_F)$ for plane strain and generalized plane stress, respectively (and similarly for $\kappa_S$).

$M_1[z, \hat{t}] := -2z\hat{t}e^{-z\hat{t}}$, (Eqn. S21)

$M_j[z, \hat{t}] := N_j[z, \hat{t}]e^{-z(1-\hat{t})} - N_j[z, -\hat{t}]e^{-z(1+\hat{t})}$, $j = 2, \ldots, 5$. (Eqn. S22)

Here: $N_2[\hat{t}] := z(1 - \hat{t})$, $N_3[\hat{t}] = N_2[\hat{t}] - 1, N_4[\hat{t}] = N_2[\hat{t}] - (\kappa_F + 1)/2$, and $N_5[\hat{t}] = N_2[\hat{t}] + (\kappa_F - 1)/2$.

In the above equations, the expression of $a_1[z]$ has been corrected from what is presented in the original Gecit paper.[33] Below, we present some of the expressions that can be used to confirm the validity of this correction.

The function $\widehat{\Pi}[\hat{x}_F, \hat{t}, z]$ is given by:

$\widehat{\Pi}[\hat{x}_F, \hat{t}, z] := \Pi[h\hat{x}_F, ht, zh^{-1}]$, (Eqn. S23)

where, Gecit defines $\Pi$ as [§] :[33]

$\Pi[x_F, t, s] :=$
$(A[s] + 2B[s] + s x_F C[s]) \cosh[s x_F] + (D[s] + 2C[s] + s x_F B[s]) \sinh[s x_F]$. (Eqn. S24)

Here, $A[s] - D[s]$ are some unknown functions that appear in the Gecit paper.[33] These functions, along with two other unknown functions $E[s]$ and $F[s]$, are determined through satisfying the boundary/continuity conditions of the elasticity problem being considered. We present the explicit forms of $A[s] - F[s]$ below:

$P_n[s] = \dfrac{1}{\kappa_F + 1} \displaystyle\int_0^c G[t]\, p_n[s, t] dt, \quad n = 1, \ldots, 5$, (Eqn. S25)

where $G[t]$ is as defined in the Gecit paper,[31] and

$$p_1[s,t] = -2ste^{-st}, \text{(Eqn. S26a)}$$

$$p_2[s,t] = s(h+t)e^{-s(h+t)} - s(h-t)e^{-s(h-t)} + \frac{\kappa_F - 1}{2}\left(e^{-s(h+t)} - e^{-s(h-t)}\right), \text{(Eqn. S26b)}$$

$$p_2[s,t] = s(h+t)e^{-s(h+t)} - s(h-t)e^{-s(h-t)} + \frac{\kappa_F - 1}{2}\left(e^{-s(h+t)} - e^{-s(h-t)}\right), \text{(Eqn. S26b)}$$

$$p_3[s,t] = -\left(s(h+t)e^{-s(h+t)} - s(h-t)e^{-s(h-t)}\right) + \frac{\kappa_F + 1}{2}\left(e^{-s(h+t)} - e^{-s(h-t)}\right), \text{(Eqn. 26c)}$$

$$p_4[s,t] = -\left(s(h+t)e^{-s(h+t)} - s(h-t)e^{-s(h-t)}\right), \text{(Eqn. S26d)}$$

$$p_5[s,t] = -\left(s(h+t)e^{-s(h+t)} - s(h-t)e^{-s(h-t)}\right) + e^{-s(h+t)} - e^{-s(h-t)}. \text{(Eqn. S26e)}$$

---

[§] The function $\Pi$ appear as function $K$ in the original paper by Gecit.[33]

Then, (simply denoting $P_n[s]$ as $P_n$ given the complexity of the expressions):

$$A[s] = P_1. \text{(Eqn. 27a)}$$

$$B[s] = \Big(2P_1(\mu_F - \mu_S)(\kappa_S\mu_F + \mu_S) + \big((1+\kappa_F)\mu_S(P_4(-1+\kappa_S)\mu_F - P_5(1+\kappa_S)\mu_F + 2P_3\mu_S) + 2hs(2P_5\kappa_S\mu_F^2 + (-P_3 + P_4 + P_5 + P_2(-1+\kappa_S)) - (P_3 - P_4 + P_5)\kappa_S)\mu_F\mu_S + 2P_2\mu_S^2)\big)\cosh[hs] - P_1(2\kappa_S\mu_F^2 + (-1+\kappa_F)(-1+\kappa_S)\mu_F\mu_S + 2\kappa_F\mu_S^2)\cosh[2hs] + \big(P_5\mu_F(-4\kappa_S\mu_F + (-1+\kappa_F + \kappa_S - \kappa_F\kappa_S + 2hs(1+\kappa_S))\mu_S) + 2hs(2P_4\kappa_S\mu_F^2 + (P_3 + P_4 - (P_3 + P_4)\kappa_S + P_2(1+\kappa_S))\mu_F\mu_S - 2P_3\mu_S^2) + \mu_S\big((2P_3 + P_4(-1+\kappa_F))(1+\kappa_S)\mu_F + 2P_2(\mu_F - \kappa_S\mu_F + (-1+\kappa_F)\mu_S)\big)\big)\sinh[hs] - P_1(1+\kappa_F)(1+\kappa_S)\mu_F\mu_S\sinh[2hs]\Big)/(2(1+2h^2s^2)\kappa_S\mu_F^2 - (1+4h^2s^2 - \kappa_F)(-1+\kappa_S)\mu_F\mu_S - (1+4h^2s^2 + \kappa_F^2)\mu_S^2 - (2\kappa_S\mu_F^2 + (-1+\kappa_F)(-1+\kappa_S)\mu_F\mu_S + 2\kappa_F\mu_S^2)\cosh[2hs] - (1+\kappa_F)(1+\kappa_S)\mu_F\mu_S\sinh[2hs]). \text{(Eqn. 27b)}$$

$$C[s] = \Big(4hP_1s(\mu_F - \mu_S)(\kappa_S\mu_F + \mu_S) - \big((1+\kappa_F)\mu_S((P_4 + P_5 + P_4\kappa_S - P_5\kappa_S)\mu_F + 2P_2\mu_S) + 2hs(2P_4\kappa_S\mu_F^2 + (P_2 + P_3 + P_4 + P_5 + (P_2 - P_3 - P_4 + P_5)\kappa_S)\mu_F\mu_S - 2P_3\mu_S^2)\big)\cosh[hs] + P_1(1+\kappa_F)(1+\kappa_S)\mu_F\mu_S\cosh[2hs] - \big(4(P_4 + hP_5s)\kappa_S\mu_F^2 + \big((P_4 + P_5)(1+2hs-\kappa_F) + (P_4 - P_5)(-1+2hs+\kappa_F)\kappa_S + 2P_2(1+hs(-1+\kappa_S)+\kappa_S) - 2P_3(-1+\kappa_S + hs(1+\kappa_S))\big)\mu_F\mu_S + 2(2hP_2s + P_3(-1+\kappa_F))\mu_S^2\big)\sinh[hs] + P_1(2\kappa_S\mu_F^2 + (-1+\kappa_F)(-1+\kappa_S)\mu_F\mu_S + 2\kappa_F\mu_S^2)\sinh[2hs]\Big)/(2(1+2h^2s^2)\kappa_S\mu_F^2 - (1+4h^2s^2 - \kappa_F)(-1+\kappa_S)\mu_F\mu_S - (1+4h^2s^2 + \kappa_F^2)\mu_S^2 - (2\kappa_S\mu_F^2 + (-1+\kappa_F)(-1+\kappa_S)\mu_F\mu_S + 2\kappa_F\mu_S^2)\cosh[2hs] - (1+\kappa_F)(1+\kappa_S)\mu_F\mu_S\sinh[2hs]). \text{(Eqn. 27c)}$$

$$D[s] = -C[s]. \text{ (Eqn. 27d)}$$

$$E[s] = \mu_F \Big( -3P_4\mu_F + 3P_5\mu_F + 6hP_4s\mu_F + 6hP_5s\mu_F - 12h^2P_4s^2\mu_F + 12h^2P_5s^2\mu_F$$
$$+ 3P_4\kappa_F\mu_F - 3P_5\kappa_F\mu_F + 6hP_4s\kappa_F\mu_F + 6hP_5s\kappa_F\mu_F$$
$$- 2P_2(1+2h^2s^2)(-3+\kappa_S)\mu_F - P_4\kappa_S\mu_F - P_5\kappa_S\mu_F - 2hP_4s\kappa_S\mu_F$$
$$+ 2hP_5s\kappa_S\mu_F - 4h^2P_4s^2\kappa_S\mu_F - 4h^2P_5s^2\kappa_S\mu_F + P_4\kappa_F\kappa_S\mu_F$$
$$+ P_5\kappa_F\kappa_S\mu_F - 2hP_4s\kappa_F\kappa_S\mu_F + 2hP_5s\kappa_F\kappa_S\mu_F$$
$$+ 2P_3(1+2h^2s^2)(3+\kappa_S)\mu_F + 2(P_4-2P_5)(1+4h^2s^2+\kappa_F^2)\mu_S$$
$$+ 4P_2(-1+\kappa_F+hs(1-4hs+\kappa_F))\mu_S$$
$$- 2P_3(1-\kappa_F+4hs(1+hs+\kappa_F))\mu_S$$
$$- 2P_1(1+\kappa_F)\Big(\mu_S + \kappa_F\mu_S - hs\big((-3+\kappa_S)\mu_F + 4\mu_S\big)\Big)\cosh[hs]$$
$$+ \Big(\big(2P_2(-3+\kappa_S) - 2P_3(3+\kappa_S)$$
$$- (-1+\kappa_F)\big(P_5(-3+\kappa_S) + P_4(3+\kappa_S)\big)\big)\mu_F$$
$$+ 2(2P_2 + P_3 - (2P_2+P_3-2P_4+4P_5)\kappa_F)\mu_S\Big)\cosh[2hs]$$
$$+ 2P_1(1+\kappa_F)\Big((-3+\kappa_S+hs(3+\kappa_S))\mu_F$$
$$- 2(-1+hs+\kappa_F)\mu_S\Big)\sinh[hs]$$
$$- (1+\kappa_F)(P_4(-3+\kappa_S)\mu_F + P_5(3+\kappa_S)\mu_F$$
$$+ 2(P_2+2P_3)\mu_S)\sinh[2hs]\Big)$$
$$/(4(1+2h^2s^2)\kappa_S\mu_F^2 - 2(1+4h^2s^2-\kappa_F)(-1+\kappa_S)\mu_F\mu_S$$
$$- 2(1+4h^2s^2+\kappa_F^2)\mu_S^2$$
$$- 2(2\kappa_S\mu_F^2 + (-1+\kappa_F)(-1+\kappa_S)\mu_F\mu_S + 2\kappa_F\mu_S^2)\cosh[2hs]$$
$$- 2(1+\kappa_F)(1+\kappa_S)\mu_F\mu_S\sinh[2hs]). \quad \text{(Eqn. S27e)}$$

$$F[s] = \mu_F \Big( -\big((2P_2+2P_3-P_4+P_5+2h(P_4+P_5)s+4h^2(P_2+P_3-P_4+P_5)s^2$$
$$+ (P_4-P_5+2h(P_4+P_5)s)\kappa_F)\mu_F\big)$$
$$+ \Big(P_2+P_3+2hP_2s(-1+2hs-\kappa_F) - P_2\kappa_F - P_3\kappa_F$$
$$+ 2hP_3s(1+2hs+\kappa_F) - (P_4-P_5)(1+4h^2s^2+\kappa_F^2)\Big)\mu_S$$
$$+ P_1(1+\kappa_F)(2hs(\mu_F-\mu_S) + \mu_S + \kappa_F\mu_S)\cosh[hs]$$
$$+ \Big(2P_2\mu_F + 2P_3\mu_F + P_2(-1+\kappa_F)\mu_S + P_3(-1+\kappa_F)\mu_S$$
$$+ (P_4-P_5)\big((-1+\kappa_F)\mu_F - 2\kappa_F\mu_S\big)\Big)\cosh[2hs]$$
$$+ (1+\kappa_F)\big(P_1((2-2hs)\mu_F + (-1+2hs+\kappa_F)\mu_S)\sinh[hs]$$
$$+ (-P_4\mu_F + P_5\mu_F + (P_2+P_3)\mu_S)\sinh[2hs]\big)\Big)$$
$$/(2(1+2h^2s^2)\kappa_S\mu_F^2 - (1+4h^2s^2-\kappa_F)(-1+\kappa_S)\mu_F\mu_S$$
$$- (1+4h^2s^2+\kappa_F^2)\mu_S^2$$
$$- (2\kappa_S\mu_F^2 + (-1+\kappa_F)(-1+\kappa_S)\mu_F\mu_S + 2\kappa_F\mu_S^2)\cosh[2hs]$$
$$- (1+\kappa_F)(1+\kappa_S)\mu_F\mu_S\sinh[2hs]). \quad \text{(Eqn. S27f)}$$

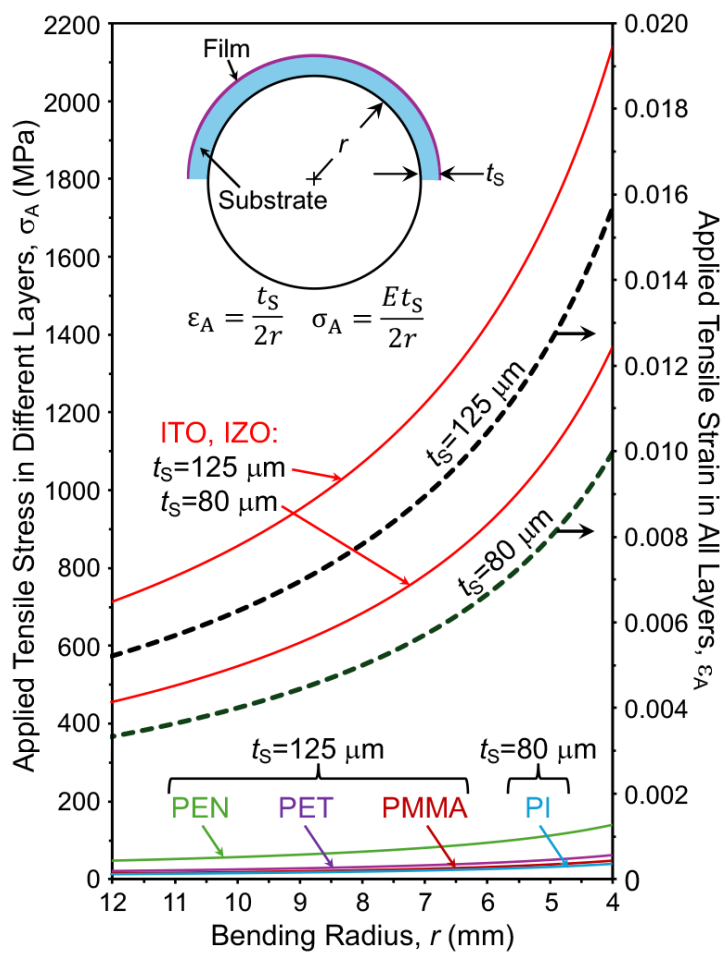

**Supplementary Fig. 1. |** Calculated applied uniaxial universal strain ($\varepsilon_A$), and the corresponding stress ($\sigma_A$) in each of the layers in bilayer film/substrate bilayers. $\varepsilon_A = t_S/2r$ and $\sigma_A = E\varepsilon_A$, where $t_S$ is the substrate thickness and $E$ is the static Young's modulus of the layer in question from Supplementary Table 1. The stress in the substrates (PEN, PET, PI) is in the region close to the top. See Supplementary Note 1 for stress/strain analyses for bending of multiple layers on a substrate.

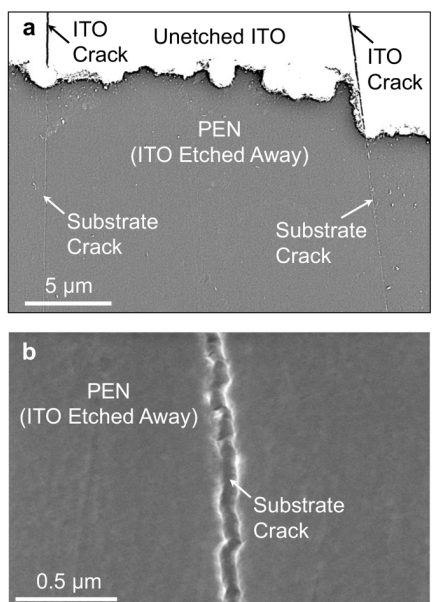

**Supplementary Fig. 2. | a, b** SEM micrographs of an ITO/PEN sheet in the bent state that was subjected to bending ($r = 7$ mm; $\varepsilon_A = 0.0089$) previously, followed by etching away of a part of the ITO film, at low magnification (**a**) and at higher magnification (**b**).

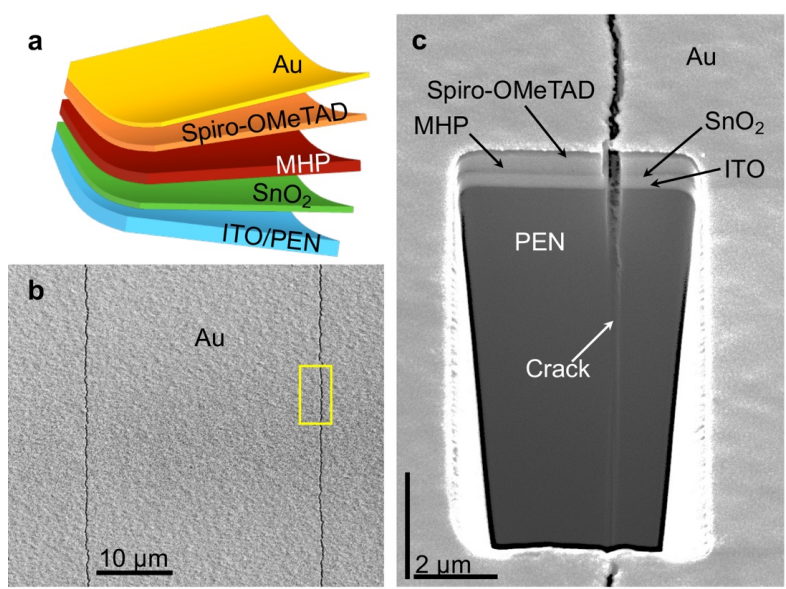

**Supplementary Fig. 3. | a,** Schematic illustration (not to scale) of the different layers in the flexible PSCs (exploded view). Here $SnO_2$ and Spiro-OMeTAD serve as the electron-transport and hole-transport, respectively, layers. Metal halide perovskite (MHP) serves as the light absorber material, and Au is the top electrode. **b-c,** SEM images of a working (19.7% efficiency) flexible PSC bent to $r = 7$ mm, in the bent state, showing top-surface view of 'channel' cracks (**b**), and FIB-cut cross-section from an area indicated by the yellow rectangle in (**b**) of a 'channel' crack revealing cracking in all the layers in the flexible PSC and the PEN substrate (**c**). (FIB-cutting was performed while in the bent state from a region similar to that indicated by the yellow rectangle in (**b**). All FIB-cut specimens were observed at a 52° forward tilt angle, hence vertical micron bar (representing depth) is longer than the horizontal one, as indicated in (**c**).

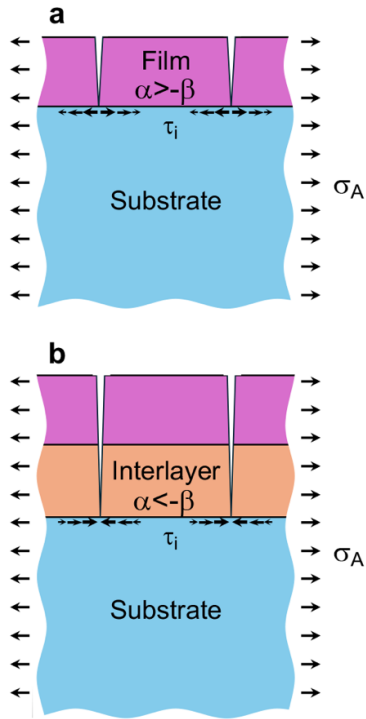

**Supplementary Fig. 4. | a, b,** Schematic illustrations (not to scale) of shear-lag concept showing generation of concentrated shear stresses ($\tau_i$) at the interface that promote substrate cracking (**a**), which are reversed in the case where low-modulus interlayer is introduced, leading to mitigation of substrate cracking (**b**).

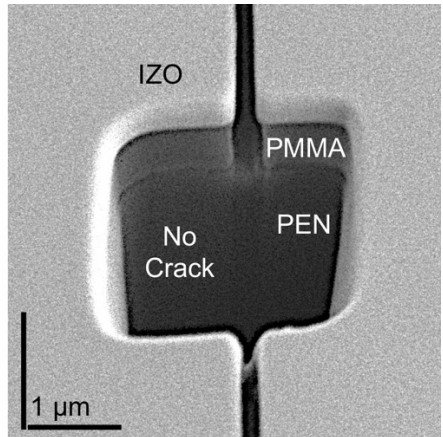

**Supplementary Fig. 5. |** SEM image of a FIB-cut cross-section of 'channel' crack in IZO/PMMA(400 nm)/PEN, bent to $r = 7$ mm, in the bent state. An ultrathin Au layer ($\sim 10$ nm) was used on top of PMMA. FIB-cuts were made while in the bent state. The FIB-cut specimen was observed at 52° forward tilt angle, hence vertical micron bars are longer than the horizontal ones, as indicated.

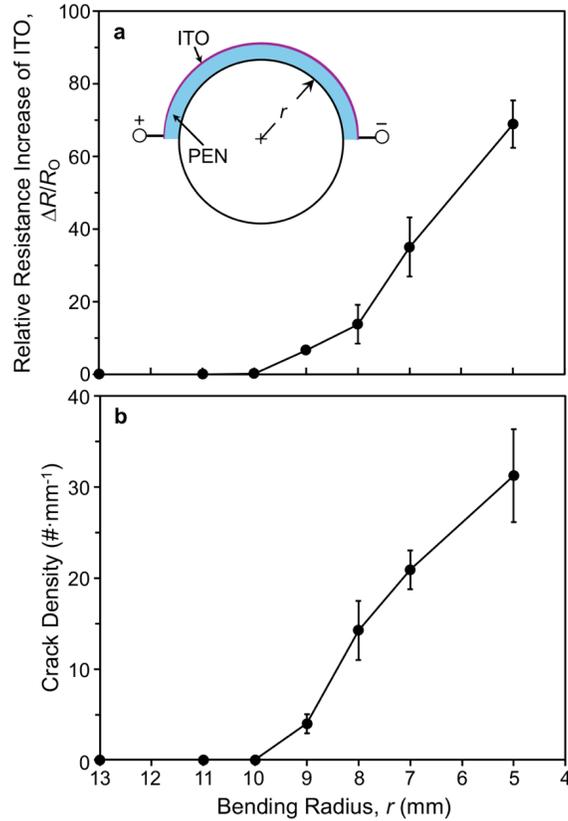

**Supplementary Fig. 6. | a,** Relative DC electrical resistance change of the ITO film in ITO/PEN sheet at various bending radii, *r*, in the bent state,. Inset: schematic illustration (not to scale) of the ITO resistance measurement in the bent state. **b,** Corresponding density of 'channel' cracks as a function of *r*.

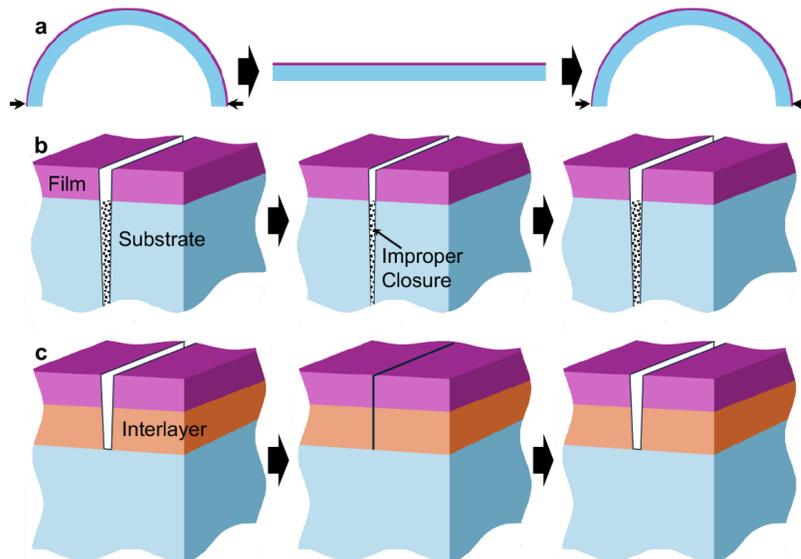

**Supplementary Fig. 7. | a,** Schematic illustration (not to scale) depicting a random cycle during the cyclic bending test: bent→flat→bent. **b, c,** Corresponding schematic illustrations (not to scale) of the hypotheses pertaining to IZO crack behavior affected by substrate cracking during cyclic bending of IZO/PEN with substrate cracking and improper closure (**b**), and IZO/PMMA(300 nm)/PEN (**c**) without substrate cracking.

## Supplementary Tables

**Supplementary Table 1. |** Thicknesses and elastic properties of the different film and substrate materials. Only static elastic properties reported in the literature are listed.

| Material | Thickness (in this work) | Young's Modulus $E^*$ (GPa) | Poisson's Ratio $\nu$ | Young's Modulus $\bar{E}^\ddagger$ (GPa) | Shear Modulus $\mu^\P$ (GPa) | Bulk Modulus $k^\S$ (GPa) | Fracture Toughness, $K_{IC}$ (MPa·m$^{0.5}$) | SEWF $w_c$ or Toughness $G_c^\dagger$ (J·m$^{-2}$) |
|---|---|---|---|---|---|---|---|---|
| PEN | ~125 μm | 9 [15] | 0.4 [15] | 10.7 | 3.2 | 15.0 | - | 55,000 [15] |
| ITO | 200 - 300 nm | 137 [18] | 0.25 [18] | 146.1 | 54.8 | 91.3 | 0.8 [18] | 5.3 [18] |
| PMMA | 200 - 600 nm | 3 [19] | 0.35 [20] | 3.4 | 1.1 | 3.3 | 2.3 [21] | - |
| IZO | ~250 nm | 130 [22] | 0.25 [22] | 138.7 | 52.0 | 86.7 | - | - |
| PET | ~125 μm | 4 [23] | 0.4 [23] | 4.8 | 1.4 | 6.7 | - | 54,000 [24] |
| PI | ~80 μm | 4 [25] | 0.34 [26] | 4.5 | 1.5 | 4.2 | - | 48,900 [24] |

$^*E$ (plane stress)  $^\ddagger \bar{E} = \dfrac{E}{(1-\nu^2)}$ (plane strain)  $^\P \mu = \dfrac{E}{\{2(1+\nu)\}}$ (isotropic)

$^\S k = \dfrac{E}{\{3(1-2\nu)\}}$ (isotropic)  $^\dagger G_c = \dfrac{K_{IC}^2\,(1-\nu^2)}{E}$ (plane strain)

**Supplementary Table 2. |** Mechanical properties (static) from Table S1 and Dundur's parameters ($\alpha$ and $\beta$) for several different film(F)/substrate(S) combinations.

| Film/Substrate | $\bar{E}_F$ (GPa) | $\bar{E}_S$ (GPa) | $\nu_F$ | $\nu_S$ | $\mu_F$ (GPa) | $\mu_S$ (GPa) | $\alpha^*$ (ref.[34]) | $\beta^\#$ (ref.[34]) |
|---|---|---|---|---|---|---|---|---|
| ITO/PEN | 146.1 | 10.7 | 0.25 | 0.4 | 54.8 | 3.2 | 0.86 | 0.13 |
| IZO/PEN | 138.7 | 10.7 | 0.25 | 0.4 | 52.0 | 3.2 | 0.86 | 0.13 |
| ITO/PET | 146.1 | 4.8 | 0.25 | 0.4 | 54.8 | 1.4 | 0.94 | 0.15 |
| IZO/PET | 138.7 | 4.8 | 0.25 | 0.4 | 52.0 | 1.4 | 0.93 | 0.15 |
| ITO/PMMA | 146.1 | 3.4 | 0.25 | 0.35 | 54.8 | 1.1 | 0.95 | 0.22 |
| IZO/PMMA | 138.7 | 3.4 | 0.25 | 0.35 | 52.0 | 1.1 | 0.95 | 0.22 |
| ITO/PI | 146.1 | 4.5 | 0.25 | 0.34 | 54.8 | 1.5 | 0.94 | 0.23 |
| IZO/PI | 138.7 | 4.5 | 0.25 | 0.34 | 52.0 | 1.5 | 0.94 | 0.22 |
| PMMA/PEN | 3.4 | 10.7 | 0.35 | 0.4 | 1.1 | 3.2 | -0.52 | -0.13 |
| PMMA/PET | 3.4 | 4.8 | 0.35 | 0.4 | 1.1 | 1.4 | -0.16 | -0.06 |
| PMMA/PI | 3.4 | 4.5 | 0.35 | 0.34 | 1.1 | 1.5 | -0.14 | -0.03 |

$^*\alpha = \dfrac{\bar{E}_F - \bar{E}_S}{\bar{E}_F + \bar{E}_S}$  $^\#\beta = \dfrac{1}{2}\dfrac{\mu_F(1-2\nu_S) - \mu_S(1-2\nu_F)}{\mu_F(1-\nu_S) + \mu_S(1-\nu_F)}$